\documentclass[a4paper]{article}
\usepackage{epsfig}

\newcommand{\tensor}{\otimes}

\newcommand{\0}{\mathbf{0}}
\newcommand{\unit}{\mathbf{1}}

\usepackage{amsfonts}
\usepackage{bbm}

\begin{document}


\title{Bayesian Probabilities and the Histories Algebra}\author{Thomas Marlow\thanks{email: pmxtm@nottingham.ac.uk}\\ \emph{School of Mathematical Sciences, University of Nottingham,}\\
\emph{UK, NG7 2RD}}

\maketitle

\begin{abstract}
We attempt a justification of a generalisation of the consistent histories programme using a notion of probability that is valid for all complete sets of history propositions.  This consists of introducing Cox's axioms of probability theory and showing that our candidate notion of probability obeys them.  We also give a generalisation of Bayes' theorem and comment upon how Bayesianism should be useful for the quantum gravity/cosmology programmes.
\end{abstract}



During this paper we will introduce a novel notion of probability within the framework of a histories theory.  Firstly, we shall introduce Cox's axioms of probability theory and then show that our proposed notion of probability obeys them.  Secondly we shall discuss the implications of such Bayesian probability assignments on the physical foundations of quantum history theories, quantum cosmology and relational gravity theories.

So, one can derive the standard rules of Bayesian probability theory from axioms that notions of probable inference should obey.  This was shown by Cox \cite{CoxBOOK} who gave two simple axioms and derived probability theory from them.  One of the axioms is that the probability that two propositions $\alpha$ and $\beta$ are both true upon a given hypothesis $I$ should functionally depend only upon the probability of one of the propositions upon the same hypothesis and the probability of the other proposition upon the same hypothesis conjoined with the presumption that the former proposition is true.  This can be written schematically as:

\begin{equation}
p(\alpha \wedge \beta \vert I) := F[p(\alpha \vert \beta I), p(\beta \vert I)].
\label{COX1}
\end{equation}

\noindent where $F$ is some function to be determined that is sufficiently well-behaved for our purposes.

Cox's other axiom is simply that the probability of the negation of a proposition upon a given hypothesis should depend only upon the probability of that proposition upon the same hypothesis.  This can similarly be written as:

\begin{equation}
p(\neg \alpha \vert I) := G[p(\alpha \vert I)].
\label{COX2}
\end{equation}

Of course, there is an implicit zeroth axiom that probabilities should be represented by real numbers.  Using such axioms Cox derived that the standard probability rules must be obeyed.

However, once one attempts to derive probability as a form of probable inference one is not wholly clear about the status of the zeroth axiom.  Certain probabilities must be considered real when they are to be interpreted as relative frequencies but, in terms of a theory of probable inference, there is no \emph{a priori} reason why probabilities should be real.  In fact, we can split up the zeroth axiom into two further axioms \cite{JaynesBOOK}.  Firstly we can presume the transitivity of probability assignments:

\begin{eqnarray}
\mbox{Axiom 0a:}& \mbox{ If } p(\alpha \vert I) > p(\beta \vert I) \mbox{ and } p(\beta \vert I) > p(\gamma \vert I) \nonumber \\ & \mbox{ then } p(\alpha \vert I) > p(\gamma \vert I)
\end{eqnarray}

\noindent where `$>$' is an ordering notion that is defined on the space we use to represent probabilities.

Secondly, we can presume what is called `universal comparability':

\begin{eqnarray}
\mbox{Axiom 0b:}& \mbox{ For all } \alpha, \beta \mbox{ we have that either } p(\alpha \vert I) > p(\beta \vert I) \nonumber \\ & \mbox{ or } p(\alpha \vert I) < p(\beta \vert I) \mbox{ or }p(\alpha \vert I) = p(\beta \vert I).
\end{eqnarray}

The combination of axioms 0a and 0b ensures that probability assignments can be real numbers.  Obviously axioms 0a and 0b are restrictions upon the type of probability space we desire for notions of probable inference.  However, we might easily not desire axiom 0b, especially in the light of quantum theory and special relativity.  There are physical reasons why we might not be able to universally compare certain propositions probabilistically.  For example, should we be able to compare two statements that involve spacelike separated regions, or that involve incompatible variables?  A form of probability that does compare such statements might involve unjustified inference.  So not only do we argue that probabilities need not be real numbers, we also argue that, in certain physical situations, they \emph{should not} be real numbers.  The reals simply might not have enough structure to represent a plausible notion of probability in certain situations.  See also \cite{Isham02} for other reasons why we might choose not to use reals for all notions of probability.

This is a thesis argued rather cogently by Youssef \cite{Youssef94,Youssef01} who argues that complex numbers and quaternions could also be consistent with Cox's two axioms (or rather axioms analogous to Cox's).  Such work attempts a derivation of the consistency of complex numbers with Cox's two axioms by presuming a distributive lattice of propositions.  One can then derive quantum mechanical features for such probability theories.  We shall not comment upon this work much more but rather we take a slightly different tack; we would like to discuss quantum history theories \cite{Grif84,Omnes88,GH90,Isham94}.

Following \cite{Isham94}, we can define a homogeneous history as an ordered tensor product of Heisenberg picture projection operators:

\begin{equation}
\alpha := \hat{\alpha}_{t_n}(t_n) \tensor \hat{\alpha}_{t_{n-1}}(t_{n-1}) \tensor ...\hat{\alpha}_{t_2}(t_2) \tensor \hat{\alpha}_{t_1}(t_1)
\label{homogeneous}
\end{equation}

\noindent such that $t_n > t_{n-1} > ... t_2 > t_1 > t_0$ and $\hat{\alpha}_{t_n}(t_n) = \hat{U}(t_n - t_0)\hat{\alpha}_{t_n} \hat{U}^{\dagger}(t_n - t_0)$ where $\hat{\alpha}_{t_n}$ is a standard Schr\"odinger picture operator and $\hat{U}$ is the standard unitary evolution operator.  The ordered set of times upon which an homogeneous history is defined is called its temporal support.  Inhomogeneous histories can then be defined using `$\vee$' or `$\neg$' operations that are naturally defined \cite{Isham94} to produce an algebra of history propositions.  There are also natural notions of disjointness and exhaustivity.  In the consistent histories programme one normally defines what is called the decoherence functional $d$ which can then be used to define relative frequencies for some complete (disjoint and exhaustive) sets of histories---these are the same relative frequencies as predicted by the von Neumann collapse hypothesis \cite{Dios04}.  Sets in which the relative frequencies are well-defined are called $d$-consistent and not all complete sets are $d$-consistent.  However, we do not wish to discuss relative frequencies \emph{per se}; we would rather discuss the more general notion of Bayesian probabilities.

One can naturally define what is called the class operator of such a history \cite{Isham94}:

\begin{equation}
C_{\alpha} := \hat{\alpha}_{t_n}(t_n) \hat{\alpha}_{t_{n-1}}(t_{n-1}) ...\hat{\alpha}_{t_2}(t_2) \hat{\alpha}_{t_1}(t_1).
\end{equation}

\noindent  Such class operators can also be defined for inhomogeneous combinations of homogeneous histories \cite{Isham94} in a natural manner.  We have the property that, for disjoint homogeneous histories that are defined over the same temporal supports, the class operators just add:

\begin{equation}
C_{\alpha \vee \beta} = \hat{\alpha}_{t_n}(t_n) ... \hat{\alpha}_{t_1}(t_1) + \hat{\beta}_{t_n}(t_n) ... \hat{\beta}_{t_1}(t_1).
\end{equation}

Instead of using the subset of the set of complete sets of history propositions that are $d$-consistent one can define a larger subset that are Linearly Positive (LP)---this larger subset was introduced by Goldstein and Page \cite{GP95} as an alternative to the $d$-consistent subset which is used in the standard consistent histories programme.  Recent work \cite{Marlow05e} has shown that there is a certain amount of consistency between a real notion of Bayesian probability and the LP formalism.  These LP probabilities seem to obey Bayesian reasoning whereas the standard notion of probability using the decoherence functional does not.  Thus we have a `good'---it seems to obey Cox's axioms---notion of probability for the LP subset of the space of history propositions.  We do not, however, have an assignment that is good for all complete sets.  All in all, it seems rather implausible that such a real probability can be found.  We might instead be able to invoke a complex notion of probability.  The decoherence functional gives complex numbers but it does not obey (\ref{COX2}) and is designed specifically with the aim of giving real probabilities.  Note that its form is derived specifically by presuming von Neumann collapse which gives a natural notion of relative frequency \cite{Anast04}. So, von Neumann collapses are designed to give something real.  Without presuming von Neumann collapse one is also released from discussing solely relative frequencies.

The most obvious complex candidate is simply:

\begin{equation}
p(\alpha \vert I) := \mbox{tr}(C_{\alpha} \rho)
\label{complexprob}
\end{equation}

\noindent where $\rho$ is the initial density matrix (defined at $t_0$) and $\alpha$ is either homogeneous as in (\ref{homogeneous}) or is an inhomogeneous proposition defined by combining homogeneous histories using the natural `$\vee$' and `$\neg$' operations \cite{Isham94}.  The real part of this candidate behave like normal probabilities for the LP subset \cite{GP95,Marlow05e}.  Also, given the natural algebra of history propositions \cite{Isham94}, (\ref{complexprob}) obeys (\ref{COX2}):

\begin{equation}
p(\neg \alpha \vert I) = \mbox{tr}((\unit - C_{\alpha})\rho) = \mbox{tr}(\rho) - \mbox{tr}(C_{\alpha} \rho) = G[p(\alpha \vert I)]
\end{equation}

\noindent where $\unit$ is the unit history proposition.

If Cox's other axiom (\ref{COX1}) is to be obeyed in situations where we have the associativity of the `$\wedge$' operation such that:

\begin{equation}
\alpha \wedge (\beta \wedge \gamma) = (\alpha \wedge \beta) \wedge \gamma = \alpha \wedge \beta \wedge \gamma
\end{equation}

\noindent then Bayes' rule should be valid (by Cox's proof \cite{CoxBOOK}).  Hence we should have that:

\begin{equation}
p(\alpha \vert \beta I) = \frac{\mbox{tr}(C_{\alpha \wedge \beta} \rho)}{\mbox{tr}(C_{\beta} \rho)}
\label{Bayes}
\end{equation}

\noindent which is well-defined as long as $\mbox{tr}(C_{\beta} \rho) \neq 0$.  Note that the `$\wedge$' is commutative for homogeneous histories defined upon the same temporal support such that $\alpha \wedge \beta = \beta \wedge \alpha$.  Hence Cox's proof of the multiplication rule also remains valid:

\begin{equation}
p(\alpha \vert \beta I) p(\beta \vert I) = p(\beta \vert \alpha I) p(\alpha \vert I).
\end{equation}

Now we must ask ourselves whether such a complex probability assignment is (a) consistent (b) unique and (c) useful?  Since it obeys Cox's axioms then it has a certain amount of consistency as a probability assignment.  Question (b) is quite hard to answer but we can certainly begin to tackle (c).  Hartle \cite{Hartle04} has recently suggested (also see \cite{Feynman87}) that virtual probabilities---real but not within $[0,1]$---might be useful as intermediate steps in any quantum analysis.  In light of axiom 0b we would prefer to go the whole hog and discuss a different space for the probabilities altogether---Hartle's virtual probabilities are real so should be rejected unless one weakens one of the probability axioms, but axiom 0b is the most plausible to weaken and would give something other than the reals; thus it seems more natural to invoke the full complex probability.  Also note the rather good analogy between the above and Youssef's work \cite{Youssef01}.  Youssef attempts to prove that complex and quarterion probabilities are compatible with Cox's axioms (or rather some axioms akin to Cox's) combined with the presumption that there exists a sublattice where the normal real probabilities are obtained.  This is a necessary feature of such a theory since axiom 0b should presumably apply to a subset of propositions, just not universally.  Such an argument adds weight to Youssef's derivation of quantum-like features---similarly for Caticha's recent pedagogical derivation of a quantum theory \cite{Catich98}.  Similarly we could ask the same question for history theories---and we have shown elsewhere \cite{Marlow05e} that there is a subset of histories where the standard Bayesian rules apply---but we already have a complex object that might obey Cox's axioms; namely (\ref{complexprob}).

If we are willing to accept a complex probability then note that the natural candidate behaves in a rather nice way.  When histories $\alpha$ and $\beta$ are homogeneous (and defined over the same temporal support) and disjoint ($\alpha \wedge \beta = \0$ where $\0$ is used to denote the null history) then we have that:

\begin{equation}
\mbox{tr}(C_{\alpha \vee \beta} \rho) = \mbox{tr}(C_{\alpha} \rho) + \mbox{tr}(C_{\beta} \rho).
\end{equation}

So, in such situations our complex probability assignment behaves in the completely standard manner:

\begin{equation}
p(\alpha \vee \beta \vert I) = p(\alpha \vert I) + p(\beta \vert I) - p(\alpha \wedge \beta \vert I).
\label{OR}
\end{equation}

Although this does not yet convince us that such an assignment is really useful conceptually, at least it behaves in a nice manner for such a large number of histories---it is valid \emph{for all complete sets of homogeneous histories}.

Note that $\alpha \wedge \beta = \beta \wedge \alpha$ for homogeneous histories defined over the same temporal support regardless of commutation issues at each time point.  In the simple case of two homogeneous histories defined over the same temporal support, where the projection operators at each time point commute, we have that:

\begin{equation}
\mbox{tr}(C_{\alpha \vee \beta} \rho) = \mbox{tr}(C_{\alpha} \rho) + \mbox{tr}(C_{\beta} \rho) - \mbox{tr}(C_{\alpha \wedge \beta} \rho).
\end{equation}

\noindent  For example, for two two-time histories in the HPO formulation $\alpha = \hat{\alpha}_{t_1} \tensor \hat{\alpha}_{t_2}$ and $\beta = \hat{\beta}_{t_1} \tensor \hat{\beta}_{t_2}$ such that $[\hat{\alpha}_{t_1}, \hat{\beta}_{t_1}] = 0$ and $[\hat{\alpha}_{t_2}, \hat{\beta}_{t_2}] = 0$ we have that:

\begin{eqnarray}
\alpha \vee \beta &=& \alpha + \beta - \hat{\alpha}_{t_1}\hat{\beta}_{t_1} \tensor \hat{\alpha}_{t_2} \hat{\beta}_{t_2} \\ &=& \alpha + \beta - (\hat{\alpha}_{t_1} \wedge \hat{\beta}_{t_1}) \tensor (\hat{\alpha}_{t_2} \wedge \hat{\beta}_{t_2}).
\end{eqnarray}

\noindent So, at least for disjoint and commuting homogeneous histories, our probability assignment behaves in the standard manner.

The major use of this `Bayesian Histories' formalism is its simplicity;  and its ease of generalisation.  What we have is a formalism that behaves as a standard probability theory for all complete sets of histories.  The probabilities just happen to be complex.  One might like to ask \emph{why} we should use complex numbers and we can see that they do not obey axiom 0b.  Youssef \cite{Youssef01} argues that, of the spaces that don't obey 0b, we should use complex or quaternion numbers because they can also obey Cox's other two axioms.  Note that the natural partial orders on complex numbers can obey axiom 0a.  He uses a distributive lattice for his argument so we have yet to complete his argument fully for the non-distributive history algebra proper.  We shall not attempt to do such a thing here.

Even given such complex probabilities, we always know that we can get standard real Bayesian probabilities out for a certain subset of the space of history propositions and, furthermore, we can get out relative frequencies by further defining $d$-consistency (note that $d$-consistency might not be the \emph{only} way for us to get relative frequencies).  The point is that we don't need all that, we know that we can get such things out in the end; for now we can just search for well-behaved Bayesian probabilities in domains not yet studied.  Hence it is the simplicity and generality of such an unreal probability programme which means it might be useful in the quantum gravity domain.  Note that what type of space of probability we use is dependent upon the space and logic of the history propositions we invoke; note also that we require notions of kinematics and dynamics before we can discuss our probability assignments (we use the Heisenberg picture).  So perhaps quantum gravity will benefit from explicitly not using standard quantum theory (as is usually the case) and we can simply search for a consistent Bayesian probability assignment for whatever propositional space one ends up deriving by other means, say using causal sets (which, presuming some form of background independence, should include both kinematical and dynamical aspects of the theory).  Youssef \cite{Youssef01} has shown that even for distributive logics one can derive many quantum mechanical features from just invoking a complex probability, so a general path is clear:  try and find a natural proposition space to be derived prior to probabilistic notions, and then invoke Bayesian reasoning to get the quantum mechanical features from the theory.

So, weirdly enough, Bayesianism might be very useful in relational theories of gravity (where it is rarely invoked; see \cite{Poulin05,Catich05} for tentative proposals in this area).  Here we would briefly like to discuss a few curious analogies between relationism and Bayesian philosophy.  There are two basic principles of relationism according to Leibniz.  Firstly there is the `principle of sufficient reason', and secondly there is the `principle of identifying the indiscernible' (see \cite{Smolin05} for an accessible discussion of relationalism).  These two principles are connected by the Bayesian `principle of \emph{in}sufficient reason' which states that if you do not have a rational reason for differentiating two statements you should assign them identical probabilities.  If you do have a rational reason for differentiating statements then you should assign different probabilities according to rational rules.  Thus, foundationally, Bayesian probability theory is wholly compatible with a relational philosophy.  In fact, these two philosophical standpoints might be identified in the future.  Bayesian probability is a way of representing such relational ideas via a probability space---the relational nature of gravity theories may help us, rather than hinder us, in searching for quantum probabilities.  Hence why we believe this Bayesian histories programme may be useful for quantum gravity theories.  We intend to investigate such a quantum gravity programme in future work.

Of course, acceptance of the above programme relies significantly upon a Bayesian view of probabilities.  So, let me now briefly discuss why such a viewpoint is useful.  Firstly note that Bayesian probabilities aren't incompatible with notions of relative frequency.  Quite the opposite;  Bayesian probabilities can incorporate most notions of relative frequencies within the literature, whereas theories of relative frequencies need to be \emph{designed} for the problem at hand \cite{JaynesBOOK}.  Bayesian probability theory is an umbrella philosophy that can incorporate many different notions of relative frequency (using notions of independence, exchangability, and maximum entropy for example).  Also, Bayesian probability is pedagogically useful because of the lack of philosophical presumptions that goes into the theory.  One can show that any notion of probable inference that obeys completely transparent axioms must behave as we expect.  Caticha has also recently shown that entropy formulae can also be derived in a particularly Bayesian way \cite{Catich03} (and references therein).  Entropy is normally considered a \emph{physical} property of systems but Caticha shows that one can sometimes consistently take an opposing view.  What better way to investigate a concept than to explicitly give the axioms by which we are allowed to consistently state it?  Especially since there might be physical cases where such axioms are violated.  This is the pedagogical power of the Bayesian programme.  Since such axioms are clearly defined, the Bayesian programme is also ripe for generalisation because certain axioms can be relaxed if there are cogent reasons for doing so.  This is not the case with relative frequencies which do not have such a clear pedagogical basis.

Often Bayesian probabilities are rejected because they are considered subjective.  This, however, is the wrong way to look at it.  It is only by considering probabilities as subjective that we can begin to understand the reasons we use the concept `probability' as we do.  Calling probability `objective', when we can never measure it directly, is not good ontology, especially since such `objective' probabilities are usually invoked as relative frequencies which in turn are defined using an entirely unphysical notion of infinite ensembles.  One can never measure `entropy' or `probability', one can only apply (either consistently or inconsistently) these concepts to the measurements we make \cite{Mana04}. `Entropy' and `probability' are forms of reasoning we \emph{use}, not things that are. We must work out \emph{why} and \emph{how} we use these concepts. Assuming that they are objective properties of systems does not help us in this enterprise as it only gives us the opportunity to accept them without thinking about why we use them in the way we do.  So, it is better to derive such concepts from a consistent set of plausible axioms.  Having to invoke some subjectivity in the notion of probability is thus not the result of the programme, it is the whole pedagogical basis of doing things in a Bayesian manner.  One must first assume that probabilities are subjective in order to then begin to work out why we ought to assign a certain probability over another.  It is also clear that these plausible axioms are not \emph{a priori} philosophical axioms; we require them to be consistent with the physics we are doing, and the physics we are doing can justify further generalisations or axioms.  So the term `subjective' should not be considered synonymous with `not physical' or `arbitrary'.

Another reason Bayesian probabilities are sometimes rejected (especially by quantum cosmologists) is because of the notion of `observers' that is often kept explicit.  Perhaps this is a misconception however, as is shown by the pervasive use of Bayesian methods in the astrophysics community.  Bayesian probability doesn't require that we have observers floating around and that we must model them---the observers need not be `in' the theory nor `outside' the system being discussed.  Bayesianism is more about what `we', as theorists, are allowed to consistently say about a theory.  There is no measurement problem as soon as one accepts that it is \emph{necessarily} `us' who are interpreting a theory.  We can either interpret a theory consistently or inconsistently; Bayesianism is an attempt to do so consistently.  Thus we do not need to invoke `observers'---one can do so for pedagogical reasons but it is not a necessary feature of Bayesian physics.  One should rather use rational `interpreters' instead of `observers' but even that is just a pedagogical notion and could still be removed from the foundations of the theory.  Interpreters are pedagogically invoked for the same reason that we need subjectivity; we must first presume that we could interpret things differently before we can begin to constrain how we ought to interpret things.  No two equals are the same.  So, such interpreters are not `passive' because they are rational but nor are they `invasive'---we don't have to assume that they (`we') are physically effecting the world around them (`us') through rationalising.

Note that implicit in our definition of probability is a notion of dynamics (we are using the Heisenberg picture).  We also explicitly have a notion of initial state $\rho$.  This may confuse some Bayesian practitioners because $\rho$ is often considered subjective and we should be able to update any state assignment given further information.  However, if we are to be discussing closed quantum systems then the dynamics \emph{and} the initial state must first be postulated \cite{Hartle05}.  They could be postulated through Bayesian reasoning, but we do not discuss such a possibility here.

Clearly we are not presenting a complete programme.  We still have to argue a direct path between weakening 0b and getting necessarily to complex numbers for the history algebra.  This would presumably involve a physical justification for not probabilistically comparing those statements that a complex probability allows us not to compare universally.  Note, however, the analogy between Bayesianism and Gleason's theorem: (\ref{COX1}) and (\ref{COX2}) are effectively non-contextuality assumptions.  Hence we remain hopeful that a uniqueness theorem may be forthcoming, perhaps analogous to the work in \cite{ILS94}.  We are also still not clear how the notion of entropy should be generalised in such domains, but we do not rule out that it also might be complex.  Note that the real part of (\ref{complexprob}) behaves like a normal Bayesian probability for the LP subset so the normal formula of Shannon entropy should also behave as normal for the LP subset \cite{Marlow05e}.

Note that we can justify the use of complex numbers in the following na\"{\i}ve manner.  For single-time propositions we have that $\hat{P}^\dagger = \hat{P}$ but for history propositions we have that $C_{\alpha}^{\dagger \dagger} = C_\alpha$.  By conjugating twice we get back to the original assignment (\emph{cf.} (\ref{COX2})).  So by reasoning in a Bayesian manner we may wish to invoke a third axiom above and beyond Cox's two:

\begin{equation}
p(M \alpha M  \vert I) := H[p(\alpha \vert I)],
\label{COX3}
\end{equation}

\noindent where $H$ is some function to be determined that is sufficiently well-behaved for our purposes, and $M$ acts on tensor product vectors so as to change the order of its entries $M(v_1 \tensor v_2 \tensor ... v_m) := (v_m \tensor v_{m-1} \tensor ... v_1)$---therefore $M \alpha M$ is the homogeneous history proposition that represents the time reversed situation (since we are in the Heisenberg picture---see (\ref{homogeneous})---this reverses both the kinematical ordering and the dynamical order \cite{SavvidTHESIS}), $M (\hat{P}_1 \tensor \hat{P}_2 \tensor ... \tensor \hat{P}_n) M:= \hat{P}_n \tensor \hat{P}_{n-1} \tensor ... \tensor \hat{P}_1$.  If we were to just change the kinematical order then we would get a different history, but by changing both we ensure a full time reversal of the history proposition.  Now, should we assign the same probability to these opposite temporal orderings?  Note that our definition of probability contains the initial state, as part of our hypothesis $I$, which hasn't been affected by the time reversal.

\begin{equation}
\mbox{tr}(C_{M \alpha M} \rho) = \mbox{tr}(C_\alpha^\dagger \rho).
\end{equation}

\noindent  So, although we have changed the order of the history proposition the initial state hasn't been changed to a final state.  The time symmetry is broken by this fact so we are rationally compelled to assign different probabilities to $\alpha$ and $M \alpha M$ upon hypothesis $I$; and if we are assign different probabilities to each then we should do so in a manner that obeys something akin to Cox's axioms so that we keep our assignments consistent---hence we invoke (\ref{COX3}).  So reversing the time ordering ensures that the history proposition is fully time reversed but we can't interpret it in a fully time reversed manner because the initial state is sequentially prior to a different projection operator, so we cannot invoke a time symmetry argument to assign equal probabilities; but nor should we order the probabilities of $p(\alpha \vert I)$ and $p(M \alpha M \vert I)$ using axiom 0b because otherwise we would be \emph{a priori} promoting a particular temporal ordering, when we can't rationally do so.  We should not irrationally presume that $p(\alpha \vert I) = p(M \alpha M \vert I)$ by some na\"{\i}ve time symmetry argument and we should ensure such assignments obey (\ref{COX3}).  Hence, perhaps, we could use complex numbers to represent probabilities.  The direction of time competes with the direction of inference.  We have used four notions of ordering: kinematical, dynamical, inferential, probabilistic---we do not want to confuse them.

Cox's two axioms suggest we must invoke probabilities that are consistent with the $\wedge$ and $\neg$ operations;  our tentative third axiom (\ref{COX3}) suggests we should also make assignments that are consistent with the $M$ operation.  We have yet to prove whether such an extra axiom necessarily forces us to use the assignment (\ref{complexprob}), although clearly (\ref{complexprob}) obeys the axioms.

Note also that the conditional complex probabilities defined using Bayes' theorem (\ref{Bayes}) behave in a nice manner; we have that, for all complete sets of homogeneous histories defined over the same temporal support $\{\alpha^i\}_{i=1}^{N_{\alpha}}$ such that $\beta$ is also homogeneous, on the same temporal support, and commutes with the $\alpha^i$:

\begin{equation}
\sum_{i=1}^{N_{\alpha}} p(\alpha^i \vert \beta I) = 1.
\end{equation}

\noindent  This is because we have a certain amount of distributivity for such histories: 

\begin{equation}
\sum_i (\alpha^i \wedge \beta) = (\sum_{i} \alpha^i) \wedge \beta.
\end{equation}

\noindent For example, for two-time homogeneous histories in the HPO formulation \cite{Isham94} we have that:

\begin{eqnarray}
(\alpha^1 + \alpha^2) \wedge \beta &=& (\hat{\alpha}_{t_2}^{1}(t_2) \tensor \hat{\alpha}_{t_1}^{1}(t_1) +  \hat{\alpha}_{t_2}^{2}(t_2) \tensor \hat{\alpha}_{t_1}^{2}(t_1)) \wedge \hat{\beta}_{t_2}(t_2) \tensor \hat{\beta}_{t_1}(t_1) \nonumber \\ &=& (\alpha^1 \wedge \beta) + (\alpha^2 \wedge \beta).
\end{eqnarray}

\noindent This property arises because history propositions are projection operators on some larger histories Hilbert space; it is a standard result that distributivity is obeyed for mutually commuting projection operators.  It passes across to the class operators such that:

\begin{equation}
\sum_{i=1}^{N_{\alpha}} \frac{\mbox{tr}(C_{\alpha^i \wedge \beta} \rho)}{\mbox{tr}(C_{\beta} \rho)} =  \frac{\mbox{tr}(C_{\unit \wedge \beta} \rho)}{\mbox{tr}(C_{\beta} \rho)} = \frac{\mbox{tr}(C_{\beta} \rho)}{\mbox{tr}(C_{\beta} \rho)} = 1.
\end{equation}

\noindent Thus updating by Bayes' theorem in this manner gives \emph{a posteriori} probabilities that behave in exactly the same manner as the \emph{a priori} ones.  For complete sets of homogeneous histories that are defined over the same temporal support, they add up to 1 and are additive (as long as the \emph{a priori} history commutes with the \emph{a posteriori} ones).  This is, of course, exactly what Bayes' theorem is about;  one takes probabilities and updates them to give something that are also good probabilities.

Another programme would be to keep probabilities real (presumably because they are to be interpreted as frequencies), but to try and give a pedagogical justification for a non-additive measure of relative frequency \cite{Sorkin95}.  Such a task would probably involve frequencies which don't converge to a single value \cite{Aerts02,Anast05}.  For frequencies that don't converge to a single value, the most natural interpretation is that we are confusing contexts somehow---this is exactly the justification given in \cite{Marlow05e} for the LP probabilities, albeit within a Bayesian framework.  However, our Bayesian analysis need not be incompatible with notions of relative frequencies as such notions (or non-convergent generalisations) should be derivable from it as in the classical case \cite{Anast05}.

So, although the presumption that probabilities are complex might initially seem patently absurd, there is a certain amount of internal consistency to the argument.  Feynman has often cogently argued for the use of `negative probabilities' in physics \cite{Feynman87} but these `negative probabilities' don't \emph{behave} like probabilities according to Cox's axioms.  The added structure provided by using complex numbers is exactly what is required in order to make a `good' notion of probability for all complete sets of homogeneous history propositions.  The real part of our complex notion can give the standard real notion of Bayesian probabilities for LP history propositions and, furthermore, can give us the standard notion of relative frequency for the $d$-consistent subset.  The berry phase is also implicit in such complex assignments.

Although not yet complete, this programme includes the two main previous quantum history theories in certain limits.  This Bayesian histories programme also suggests where we should look in order to find a physical justification for quantum history theories; namely we point at axiom 0b (and perhaps the invocation of a third axiom).  It is also foundationally compatible with a relational philosophy.

\section*{Acknowledgements}

We thank EPSRC for funding this work.


\begin{thebibliography}{99}

\bibitem{CoxBOOK} R. T. Cox, \emph{The Algebra of Probable Inference} (The Johns Hopkins University Press, 1961).

\bibitem{JaynesBOOK} E. T. Jaynes, \emph{Probability Theory: The Logic of Science}, (Cambridge University Press, 2003).

\bibitem{Isham02} C. J. Isham, ``Some Reflections on the Status of Conventional Quantum Theory when Applied to Quantum Gravity'' in \emph{Proceedings of the Conference in Honour of Stephen Hawking's birthday}, Ed., G. Gibbons (Cambridge University Press, 2003) preprint: {\tt quant-ph/0206090 v1}.

\bibitem{Youssef94} S. Youssef, ``Quantum Mechanics as Complex Probability Theory'' \emph{Mod. Phys. Lett A} \textbf{9} (1994) 2571.

\bibitem{Youssef01} S. Youssef, ``Physics with exotic probability theory'' (2001) preprint:  {\tt hep-th/0110253 v2}.

\bibitem{Grif84} R. B. Griffiths, ``Consistent history propositions and the Interpretation of Quantum Mechanics'' \emph{J. Stat. Phys.} \textbf{36} (1984) 219-273.

\bibitem{Omnes88} R. Omn\'es, ``Logical reformulation of quantum mechanics. I. Foundations'' \emph{J. Stat. Phys.} \textbf{53} (1988) 933-955.

\bibitem{GH90} M. Gell-Mann and J. Hartle, ``Quantum Mechanics in the light of quantum cosmology'' in \emph{Proceedings of the Third International Symposium on the Foundations of Quantum Mechanics in the Light of New Technology} (Physical Society of Japan, Tokyo, Japan, 1990) 321-343.

\bibitem{Isham94} C. J. Isham, ``Quantum logic and the history propositions approach to quantum theory'' \emph{J. Math. Phys.} \textbf{35} (1994) 2157-2185, preprint: {\tt gr-qc/9308006}.

\bibitem{Dios04} L. Di\'{o}si, ``Anomalies of weakened decoherence criteria for quantum histories'' \emph{Phys. Rev. Lett.} \textbf{92} (2004) 170401, preprint: {\tt quant-ph/0310181 v1}.

\bibitem{Marlow05e} T. Marlow, ``A Bayesian Account of Quantum Histories'' to be published in \emph{Ann. Phys.} (2006) preprint:  {\tt quant-ph/0509149}.

\bibitem{GP95} S. Goldstein, D. N. Page ``Linearly Positive history propositions: Probabilities for a Robust Family of Sequences of Quantum Events'' \emph{Phys. Rev. Lett.} \textbf{74} (1995) 3715-3719,  preprint: {\tt gr-qc/9403055}.

\bibitem{Anast04} C. Anastopoulos, ``On the relation between quantum mechanical probabilities and event frequencies'' \emph{Ann. Phys.} \textbf{313} (2004) 368, preprint: \\{\tt quant-ph/0403207}.

\bibitem{Hartle04} J. B. Hartle, ``Linear Positivity and Virtual Probability'' \emph{Phys. Rev. A} \textbf{70} (2004) 022104, preprint: {\tt quant-ph/0401108}.

\bibitem{Feynman87} R. P. Feynman, ``Negative Probabilities'' in \emph{Quantum Implications}, Eds., B.J. Hiley and F.David Peat (Routledge and Kegan Paul, 1987) 235.

\bibitem{Catich98} A. Caticha, ``Consistency, Amplitudes and Probabilities in Quantum Theory'', \emph{Phys. Rev. A} \textbf{57} (1998) 1572, preprint: {\tt quant-ph/9804012}.

\bibitem{Poulin05} D. Poulin, ``Toy Model for a Relational Formulation of Quantum Theory'' (2005) preprint:  {\tt quant-ph/0505081 v2}.

\bibitem{Catich05} A. Caticha, ``The Information Geometry of Space and Time'' (2005) preprint:  {\tt gr-qc/0508108 v1}.

\bibitem{Smolin05} L. Smolin, ``The case for background independence'' (2005) preprint: {\tt hep-th/0507235}.

\bibitem{Catich03} A. Caticha, ``Relative Entropy and Inductive Inference''  \emph{AIP Conference Proceedings} \textbf{707} (2004) 75-96, preprint: {\tt physics/0311093}.

\bibitem{Mana04} P. G. L. Mana, ``Consistency of the Shannon entropy in quantum experiments'' \emph{Phys. Rev. A} \textbf{69} (2004) 062108, preprint: {\tt quant-ph/0302049 v5}.

\bibitem{Hartle05} J. B. Hartle, ``Excess Baggage'' in \emph{Elementary Particles and the Universe}, Ed., J. Schwarz (Cambridge University Press, 1991) updated preprint: {\tt gr-qc/0508001}.

\bibitem{ILS94}  C. J. Isham, N. Linden, S. Schreckenberg ``The Classification of Decoherence Functionals: An Analogue of Gleason's Theorem'' \emph{Jour. Math. Phys.} \textbf{35} (1994) 6360-6370, preprint: {\tt gr-qc/9406015}.

\bibitem{SavvidTHESIS} K. Savvidou, ``Continuous time and consistent histories'' \emph{Ph.D. Thesis} (The Blackett Laboratory, Imperial College, London, 1999) gr-qc/9912076.

\bibitem{Sorkin95}  R. D. Sorkin, ``Quantum Measure Theory and its Interpretation'' in \emph{Quantum Classical Correspondence: Proceedings of the 4th Drexel Symposium on Quantum Nonintegrability} Eds., D. H. Feng and B.L Hu (International Press, Cambridge Mass. 1997) preprint: {\tt gr-qc/9507057}.

\bibitem{Aerts02} D. Aerts, ``Reality and Probability: Introducing a new Type of Probability Calculus'' in \emph{Probing the Structure of Quantum Mechanics: Nonlocality, Computation and Axiomatics}, D. Aerts, M. Czachor and T. Durt eds. (World Scientific, Singapore, 2002) preprint: {\tt quant-ph/0205165}.

\bibitem{Anast05} C. Anastopoulos, ``Classical vs Quantum Probability in Sequential Measurements'' (2005) preprint: {\tt quant-ph/0509019}.

All preprints refer to the http://axiv.org/ website

\end{thebibliography}
\end{document}